\documentstyle[12pt]{article}
\newcommand{\beq}{\begin{equation}}
\newcommand{\eeq}{\end{equation}}

\newcommand{\beqs}{\begin{eqnarray}}
\newcommand{\eeqs}{\end{eqnarray}}
\newcommand{\eqn}[1]{~(\ref{#1})}

\newcommand{\half}{\frac{1}{2}}
\newcommand{\eps}{\epsilon}

\renewcommand{\Im}{\;{\rm Im}\;}
\renewcommand{\Re}{\;{\rm Re}\;}
\begin{document}

\def\m#1{{$#1$}}
\def\p{{ p}}
\def\q{{ q}}
\def\k{{ k}}
\def\tr{\;{\rm tr}\;}
\def\hi{{\cal H}}
\def\ignore#1{{}}

\begin{center}{\Large A Dispersion Relation  for the Density of States \\ With Application to the Casimir Effect}
\end{center}
\begin{center}
{S. G. Rajeev\\
 Department of Physics and Astronomy\\
Department of Mathematics\\
University of Rochester\\
Rochester,NY 14627,USA }
\end{center}

\centerline{\bf Abstract}

The trace  of a function of a Schr\"odinger operator minus the same for the Laplacian can be expressed  in terms of the determinant of its scattering matrix. The naive formula for this  determinant is  divergent. Using a dispersion relation, we find another expression for it which is convergent, but needs one piece of  information  beyond   the scattering matrix. Except for this `anomaly', we can  express the Casimir energy of a compact body in terms of its optical scattering matrix, without assuming any rotational symmetry for its shape.

\newpage
\section{Introduction}

The kinetic theory of gases in classical physics predicts that every
surface in a gas is bombarded by molecules. The recoil
of these molecules  exert a pressure on the surface.  A spectacular
demonstration of this prediction would be to evacuate the air inside
an aluminium can: there will be a net inward pressure that will cause
the can to collapse.

In the quantum theory of fields, there is an analogous pressure on
every surface that can scatter light\cite{Casimir,CasimirReview}.  Even  in the vacuum there
are `virtual photons' which describe the quantum fluctuations of the
electromagnetic field in its ground state. These virtual photons are
scattered by any medium that can interact with light; this scattering
exerts a force on the medium. Although quite small in magnitude, it has
been measured experimentally\cite{CasimirExp}.

In the original calculation of the Casimir force only simple shapes such
as a sphere, or a rectangular slab were  studied. Moreover, the medium was   assumed to have ideal properties, such as  perfect conductivity. Since that time,  new methods have been developed, which allow the calculation of the Casimir energy for more general shapes. Of particular interest to us  are the spectral methods\cite{MITGroup,Graham,Wirzba, Dreyfus,EmigJaffe} . (Methods that allow realistic   computation of Casimir energy  of  micro-electromechanical  or MEM devices  have been developed recently\cite{SharpConductors}; but we do not use them in this paper.)
  The basic idea behind these methods  is a relationship (often called Levinsion's theorem or    Krein's  trace formula)  between the optical scattering matrix \m{S} and the density of states:
\beq
 \rho(k)={1\over 2\pi i}{d\over dk}\tr\log  S, 
 \eeq
 The Casimir energy  is the  sum  over frequencies weighted by \m{\rho(k)}. 

The physical argument that the Casimir force is due to the reflection of virtual photons suggests that it should be possible to express it entirely in terms of the probability of reflection.  Unless the momentum of the photon changes during scattering, it should not contribute to the Casimir force.

 In this paper we will show that the dispersion relations of scattering theory allow an answer for the density of states  in terms of the reflection probability alone, in the one dimensional case. In the three dimensional case, there is a potential logarithmic divergence in  \m{\tr\log S} .  We show how to  remove this  by using a dispersion relation, without assuming rotational symmetry or using perturbation theory as in previous treatments \cite{SumRules}.   An outcome is  an `anomaly'  term in the density of states proportional  to the integral of the potential. Contrary to the physical picture in terms of scattering of virtual photons, there is this one contribution to the Casimir effect that cannot be expressed in terms of the scattering matrix alone.

Much of our detailed analysis will be only carried out for the technically
simpler case of a scalar field. As in traditional optical scattering
theory, this scalar model  gives a reasonable picture of the essential
phenomenon. It is straightforward although technically involved to
extend our analysis to include polarization effects; i.e., scattering
of vectorial fields. Moreover, we will ignore the effects of absorption of light: the effect of absorption on virtual photons needs a deeper analysis than we can provide at the moment.  Also, we will consider only optical media that are time independent. For a technical reason, we will further  assume that the scatterer is parity invariant; although it does not have to be symmetric otherwise.

We aim  to give a more or less self-contained description, taking off from the discussion of scattering in standard textbooks. The erudite reader could skip ahead to sections 3.1,3.2 and 5. No pretense at mathematical rigor is made. However, there will be an occasion for us to be careful about the definition of an infinite determinant (of the scattering operator) to avoid a logarithmic divergence.

\section{ Scattering of Light}

The propagation of light of(Ref. \cite{ElecContMed})
wave-number   \m{k} in a dielectric medium is  described by the equations
\beq
{\bf \nabla}\times {\bf E}=ik {\bf H},\quad
		{\bf \nabla}\times {\bf H}=-ik \eps(x,k){\bf E}.
\eeq

Here, \m{\eps(x,k)} is the dielectric `constant' which may in fact depend on
position and on the wave-number \m{k}. (In other words, \m{\eps(x,k)}
is the square of the refractive index.) \m{\bf E} and \m{\bf H} are
complex vector valued functions describing the amplitude of the
electromagnetic wave.

Eliminating \m{\bf H}, we get
\beq
	{\bf \nabla}\times {\bf \nabla}\times {\bf
E}(x,k)=k^2\eps(x,k){\bf E}.
\eeq
In other words,
\beq
	{\bf \nabla}\times {\bf \nabla}\times {\bf
E}(x,k)+V(x){\bf E}=k^2{\bf E}
\eeq
where the `effective potential' \cite{Optics} is
\beq
	V(x)=k^2[1-\eps(x,k)].
\eeq

There is an analogy between the above equation describing  the classical
scattering of light and the quantum mechanical scattering of a
particle by a potential. The Schr\"odinger equation (in natural units
\m{\hbar=2m=1}) is
\beq
	-\nabla^2\psi+V(x)\psi=k^2\psi.
\eeq
The essential difference is that the vector Laplacian \m{{\bf \nabla}\times
{\bf \nabla}\times {\bf E}} is replaced by the scalar Laplacian
\m{\nabla^2}.
The potential \m{V(x)} is related to the refractive index as in the
formula  above. If the refractive index is less  than one,  the
`potential' is positive; this is the case for large enough \m{k}.

Thus it will be useful to use the language of quantum mechanical
 scattering theory, although of course  the scattering we 
are talking about is just the classical scattering of light waves.
But some differences must be kept in mind. For
example, the `potential' \m{V(x,k)} will depend on wavenumber, a fact
that has no simple meaning in quantum mechanics. The wave-number  dependence of the scattering matrix will be affected by this
 dispersion of the light waves. Our
 final formulas only involve the scattering matrix and thus take into
 account of this effect. But we will often not explicitly display the
 dependence of the potential on the wavenumber.

\section{ Waves in \m{1+1} Dimensions}

We will start our discussion in the simplest possible case, the scalar
 theory in one space and one time. We will later generalize the ideas
 to three dimensions and to vector waves, but it is pedagogically
 useful to start with the simplest case.

Often we are  interested in deriving a formula for the sum over a
function \m{\phi} of the eigenvalues of a  hamiltonian operator \m{H}.
This can be
thought of as the trace of an  operator \m{\tr \phi(\surd H)}.
 For example, the Casimir energy of a  massless scalar field
  in the
presence of an external potential  is \m{\tr \surd
 H}. Another example is the thermodynamic partition function,
\m{\tr e^{-\beta H}}. More
precisely we will be interested in the difference between this trace
and its value for the free hamiltonian.

With this in mind,let
\beq
	H=-{d^2\over dx^2}+V(x),\quad H_0=-{d^2\over dx^2}
\eeq
We can think of \m{H} be as the Schr\"odinger operator of a one dimensional
quantum mechanical system. The potential \m{V(x)} is positive, smooth and
vanishes faster than any power of \m{x} at infinity. Thus, \m{H\geq
0}.

 Let \m{\phi(k)} be a smooth function on \m{[0,\infty)} which vanishes at
 infinity faster than  \m{k^{-1}}. We are going to derive a
 formula for the quantity
\beq
	\bar{\tr}\phi(\surd H)=\tr [\phi(\surd H)-\phi(\surd H_0)].
\eeq
We assume, for now, that \m{\phi(k)} vanishes  for large \m{k} : postponing  the study of
 ultraviolet divergences.
 We might expect
 that there is an integral representation
\beq
	\bar{\tr}\phi(\surd H)=\int_0^\infty \phi(k)\rho(k)dk.
\eeq
We will get a formula for the `spectral density' \m{\rho(k)} in terms
 of the scattering matrix. We can
 get this directly from the formula for the resolvent in terms  of the
 Jost functions\cite{Graham,Wirzba}, but we will give a more physical but less rigorous  argument.

Instead, we will look at the continuous spectrum as the limit of  an
eigenvalue problem.
Imagine we have enclosed  the whole system in a box of size
\m{2L}: we require that the eigenfunctions of \m{H} and \m{H_0} vanish
at \m{x=\pm L}. Then we have a discrete number of allowed values of
momentum, \m{k_n(L)}. For example when \m{V(x)=0}, they are
\m{\pi n\over 2 L}. Thus
\beq
	\bar{\tr}\phi(\surd H)=\sum_n\bigg[\phi\big(k_n(L)\big)-\phi\big({\pi
n\over  2L}\big)\bigg]
\eeq
This sum will become an integral in the limit as \m{L\to \infty}.
Even when \m{L} is
kept finite, we can assume that it is  much larger than the range of
the potential \m{V(x)}, since we are only interested in the limit
\m{L\to \infty}.

At infinity, the solutions of the Schr\"odinger  equation \cite{LandauQM}
\beq
	-u''+V(x)u(x)=k^2u(x)
\eeq
tend to a sum of plane
waves:
\beqs
	u(x)&\to&Ae^{ikx}+Be^{-ikx}\; {\rm for}\; x\to -\infty\cr
	    &\to&Ce^{ikx}+De^{-ikx}\; {\rm for}\; x\to \infty.
\eeqs
Thus \m{A,D} are the amplitudes of the incoming waves and \m{C,B}
those of the outgoing waves. Conservation of probability	
gives
\beq
	|A|^2+|D|^2=|C|^2+|B|^2.
\eeq
Since the Schr\"odinger equation is a second order  differential equation,
there are two  constants of integration in its general solution. Let
us choose them to be \m{A} and \m{D}, the amplitudes of the incoming waves. Then
the amplitudes of the outgoing waves are determined once we solve the
differential equation:
\beq
	\pmatrix{C\cr B}=S(k)\pmatrix{A\cr D}.
\eeq
The matrix \m{S(k)} is unitary from the conservation of probability:
it is the scattering matrix associated to the potential \m{V}.

Often we describe the scattering in terms of reflection and
transmission coefficients. They are defined by the special choice
\m{A=1,D=0} corresponding to waves incident from the left:
\beqs
	u(x)&\to&e^{ikx}+R(k)e^{-ikx}\; {\rm for}\; x\to -\infty\cr
	    &\to&T(k)e^{ikx}\; {\rm for}\; x\to \infty.
\eeqs
Taking complex conjugate we have another solution,
\beqs
	u^*(x)&\to&e^{-ikx}+R^*(k)e^{ikx}\; {\rm for}\; x\to -\infty\cr
	    &\to&T^*(k)e^{-ikx}\; {\rm for}\; x\to \infty.
\eeqs
Taking linear combinations,  gives the solution describing waves
incident from the right:
\beqs
	u^*(x)-R^*(k)u(x)\over T^*(k)&\to& {1-|R(k)|^2\over T^*(k)}e^{-ikx}\;  {\rm for}\; x\to -\infty\cr
	    &\to&-{R^*(k)T(k)\over T^*(k)} e^{ikx}+e^{-ikx}\; {\rm for}\; x\to \infty.
\eeqs
Thus (recalling that \m{|R(k)|^2+|T(k)|^2=1}),
\beq
	S(k)=\pmatrix{T(k)&-R^*(k){T(k)\over T^*(k)}\cr
			R(k)&T(k)}
\eeq
We might check  explicitly that this matrix is unitary.

Being unitary, there is a basis in which \m{S(k)} is diagonal, with
eigenvalues\m{\ e^{2i\eta_{1,2}(k)}}. The real valued functions
\m{\eta_{1,2}(k)} are the `phase shifts' of the scattering problems.

Now let us return to the eigenvalue problem
\beq
	-u''+V(x)u(x)=k^2u(x),\quad u(-L)=u(L)=0
\eeq
Since  \m{L} is large compared to the range of \m{V(x)}, we can use
the asymptotic forms above in the boundary conditions:
\beq
	Ae^{-ikL}+Be^{ikL}=0=Ce^{ikL}+De^{-ikL}.
\eeq
Solving for \m{B} and \m{C},
\beq
	B=-Ae^{-2ikL},\quad C=-De^{-2ikL}.
\eeq
Thus the momenta \m{k} are determined by the equation
\beq
	S(k)\pmatrix{A\cr D}=-e^{-2ikL}\pmatrix{D\cr A}.
\eeq
So far we are in arbitrary basis in the two dimensional space of
solutions. Now let us choose this basis to the one that digaonalizes
\m{S(k)}:
\beq
	e^{2i\eta_1(k)}A=-e^{-2ikL}D,\quad
e^{2i\eta_2(k)}D=-e^{-2ikL}A.
\eeq
Eliminating \m{D},
\beq
	e^{2i[\eta_1(k)+\eta_2(k)+2kL]}=1.
\eeq
This is the transcendental equation  for the allowed wavenumbers,
\beq
	k={\pi n \over 2 L}-{\eta_1(k)+\eta_2(k)\over 2L},\;{\rm
for}\; n=\cdots -2,-1,0,1,2,\cdots
\eeq
Since \m{L} is  very large in the limit we are interested
in,  the second term is  a small correction. It
is enough to
 solve the equation approximately by iterating it once:
\beq
	k_n(L)={\pi n \over 2L}-{\eta_1\big({\pi n\over 2L}\big)+\eta_2\big({\pi
n\over 2L}\big)\over 2L}
\eeq
We can now reexpress the sum over phase shifts  in terms of the Scattering matrix:
\beq
2i[\eta_1(k)+\eta_2(k)]=\log\det S(k).
\eeq
For later use we also note that,
\beq
	\det S(k)=T^2(k)+{|R(k)|^2T(k)\over T^*(K)}={T(k)\over T^*(k)}
\eeq
so that
\beq
	\log\det S(k)=2i\arg T(k).
\eeq
Thus,
\beq
k_n(L)-{\pi n \over 2L}=-{\log\det S\big({\pi n\over  2L}\big)\over 4iL}
\eeq

Now we are ready to evaluate the sum over momenta in the limit as
\m{L\to \infty}.
\beqs
	\bar{\tr}\phi(\surd H)&=&\sum_n\bigg[\phi\big(k_n(L)\big)-\phi\big({\pi
n\over 2L}\big)\bigg]\cr
&=&- \sum_n\phi'\big({\pi n\over 2L}\big)
                     {\log\det S\big({\pi n\over  2L}\big)\over 4iL}
\eeqs
This becomes an integral as \m{L\to \infty}:
\beq
	\bar{\tr}\phi(\surd H)=-\int_0^\infty{dk\over  2\pi
i}{d\phi(k)\over dk}\log\det S(k)
\eeq
Similar   trace formulas have been derived by many other methods\cite{Graham, Wirzba,SumRules}.

 If the potential is independent of frequency, the integral will converge if
\m{\phi(k)} falls off faster than \m{k^{-1}} for large \m{k} and
remains finite for small \m{k}. (This is because \m{T(k)\sim 1+{1\over
2ik}\int V(x)dx} for large \m{k} from the Born approximation; more on
this later.) If \m{\phi(k)} does not fall off at large \m{k}, the integral can still converge if \m{V(x,k)} tends to zero for large \m{|k|}, so that \m{S(k)} tends to unity faster also.

 Even if  the potential depends on the wave-number, it will still determine
a unitary scattering matrix \m{S(k)} for each  \m{k}. We can still ask
for the sum  over the allowed values of wave-number for finite \m{L} as
before; it is determined by the same formula  in terms of the
scattering matrix. Indeed, the whole argument goes through without any
change.

\subsection{Dispersion Relation}

We will now express the density of states in terms of the reflection
coefficient. The force due to the scattering of a virtual particle
ought to be dependent  on the probability of reflection: if the
particle is transmitted, its momentum does not change and hence it
should exert no force. Thus it is physically unsatisfactory that we
have a formula in terms of the argument of the transmission amplitude
rather than the magnitude of the reflection amplitude.

But we now remember the basic fact that \m{T(k)} is analytic in the
upper half plane. There are no poles  since there are no bound
states. Moreover, \m{T(k)} never has zeros in the upper half
plane. Hence \m{\log T(k)} is also analytic in the upper half
plane. There is then a dispersion relation \cite{LandauQM}  between its real and
imaginary parts (Hilbert transform) :
\beq
	\Im \log T(k)={1\over \pi}{\cal P}\int {\Re \log T(k')\over
k'-k}dk'.
\eeq
But
\beq
	\Re \log T(k)=\half \log|T(k)|^2=\half \log[1-|R(k)|^2].
\eeq
Thus
\beq
	\arg \det S(k)= 2\arg T(k)={\cal P}
	\int {\log[1-|R(k')|^2]\over k'-k}{dk'\over\pi}.
\eeq
\ignore{\bf Work out the special case of Casimir energy integral in terms of reflection coefficients.
}
\subsection{The Trace  In Terms of  the Reflection Coefficient}
 Combining the above results we get
\beqs
	\bar{\tr}\phi\left(\surd H \right)&=&-\int_0^\infty{dk\over 2\pi
}{d\phi(k)\over dk}\arg\det S(k)\\
&=&{\cal P}\int_0^\infty{dk\over 2\pi}\int_{-\infty}^{\infty}{dk'\over\pi}
{d\phi(k)\over dk}
	 {\log[1-|R(k')|^2]\over k-k'}\label{trphi1d}
\eeqs
Assuming parity invariance \m{|R(k)|=|R(-k)|},
\beqs
	\bar{\tr}\phi\left(\surd H \right)&=&{\cal P}\int_0^\infty{dk\over \pi}{dk'\over\pi}
{d\phi(k)\over dk}
	 {k\log[1-|R(k')|^2]\over k^2-k'^2}
\eeqs
We can expand the log in a power series to get a `multiple reflection expansion' for this quantity.

When \m{\phi(k)=k} this can be simplified to
\beqs
	\bar{\tr}\surd H &=&{1\over2}\int_0^\infty{dk\over \pi}{dk'\over\pi}
 \ \ {  k\log[1-|R(k')|^2]  -k'\log[1-|R(k)|^2]      \over k^2-k'^2}
\eeqs
with  a non-singular integrand.
\ignore{\subsection{ Examples: Leave This Out}
Perhaps the simplest example is that of an infinitesimally thin
potential barrier, described by a delta function:
\beq
	V(x)=g\delta(x),\quad g>0.
\eeq
Standard quantum mechanics gives the transmission amplitude
\beq
	T(k)={2ik\over 2ik-g}=\big[1+{ig\over 2k}\big]^{-1}.
\eeq
Thus
\beq	
	{1\over 2\pi i}\log\det S(k)={1\over \pi}\arg T(k)=
-{1\over \pi}\arctan({g\over 2k}).
\eeq
It is as though  this thin barrier has  one half state less than the
vacuum,  when we sum  over all the values of wavenumber.
}

\section{ Waves in  \m{3+1} Dimensions}

Consider the operator  \m{H=- \nabla^2+V(x)}, on \m{L^2(R^3)}
where the `potential' \m{V(x)}
is a real function  on \m{R^3} that vanishes at infinity. (The
potential also can  depend on the wave-number: but again we don't display this explicitly.) We will
also assume for simplicity that
that \m{H}  has positive continuous spectrum ; i.e., that it has no `bound
states'.

We will again be interested in the   the difference
\beq
	\bar{{\rm tr}}(\phi(\surd H))=\tr\phi(\surd H)-\tr\phi(\sqrt{H_0})
\eeq
of the trace from that with the free hamiltonian:
\beq
	H_0=-\nabla^2.
\eeq
The function \m{\phi(k)} will, to begin with,  be assumed to be smooth and vanish at
infinity faster than \m{k^{-3}}: this is necessary to avoid
`ultra-violet' divergences. Later on we will allow for \m{\phi(k)=k} and check that the integrals converge when the wave number dependence of the potential is take into account.
Again we will derive a formula that involves the scattering matrix.

 Since the potential vanishes at infinity,  there is a  solution to
the  equation
\beq
	[-\nabla^2+V(x)]\psi(x)=k^2\psi(x)
\eeq
that tend to   a sum of a plane wave and an outgoing spherical wave\cite{LandauQM}:
\beq
	\psi(r,{\bf n}')\to  e^{ikr{\bf n}\cdot{\bf n}'}+f(k,{\bf n},{\bf
n}'){e^{ikr}\over r}, \;\quad {\rm as}\; r\to \infty
\eeq
 Here, \m{\bf n} is the direction of  the incoming plane wave and
\m{{\bf n}'} the direction along which we let the argument of the
wavefunction go to infinity.
The function \m{f(k,{\bf n},{\bf n}')} is
the scattering amplitude.

Since there are no bound states, the general solution  can be written
as a superposition of such scattering
solutions, \cite{LandauQM} weighted by a function \m{F({\bf n})} of the
direction of incidence. Such a general solution will have the
asymptotic behavior at spatial infinity,
\beq
	\psi_F(r,{\bf n}')\sim \int F({\bf n})e^{ikr{\bf n}\cdot{\bf n}'}d\Omega_{\bf n}+
{e^{ikr}\over r}\int F({\bf n})f(k,{\bf n},{\bf n}')d\Omega_{{\bf n}}
\eeq
As \m{r\to \infty} the first integral can be evaluated by the method
of steepest descents, to get
\beq
2\pi iF(-{\bf n}'){e^{-ikr}\over kr}-2\pi iF({\bf n}'){e^{ikr}\over kr}.
\eeq
Thus we can write
\beq
	\psi_F(r,{\bf n}')\sim {2\pi i \over k}\bigg\{{e^{-ikr}\over r}F(-{\bf
n}')-{e^{ikr}\over r}[1+2ik\hat{f}(k) F]({\bf n}')\bigg\}
\eeq
where  \m{\hat{f}(k)} is the operator on \m{L^2(S^2)}:
\beq
	\hat{f}(k) F({\bf n}')={1\over 4\pi}\int f(k,{\bf n},{\bf
n}')F({\bf n})d\Omega_{\bf n}.
\eeq
The quantity
\beq
	\hat S(k)=1+2i k\hat f(k)
\eeq
is the scattering operator. Conservation of probability requires it to
be unitary.

Being unitary this operator can be diagonalized on \m{L^2(S^2)}. If
\m{H} is spherically symmetric, the eigenfunctions of \m{\hat S} are
the spherical harmonics:
\beq
	\hat S(k)Y_{lm}({\bf n})=e^{2i\eta_l(k)}Y_{lm}({\bf n}).
\eeq
The eigenvalues are determined by the `phase-shifts' \m{\eta_l(k)} in
each angular momentum sector. Even if \m{H} is not spherically
symmetric, there will be some
spectrum of `phase shifts' \m{\eta_a(k)}:
\beq
	\hat S(k)\chi_a(k,{\bf n})=e^{2i\eta_a(k)}\chi_a(k,{\bf n}).
\eeq
(We will assume for now that this spectrum of \m{\hat S} is discrete; i.e.,
that the label \m{a} takes values in a countable set.)

Let us first imagine that our whole system is enclosed in a spherical
box of large radius \m{R} with  the wavefunction required to  vanish on this
surface. (The exact shape doesn't matter in the limit \m{R\to \infty},
which is all we are interested in, so we might as well assume it is
spherical.) For finite \m{R} there is a discrete set of
allowed  values for \m{k}, say \m{k_{na}(R)}. They are fixed by the
condition
\beq 	
	\psi(R,{\bf n'})=0.
\eeq
When \m{R} is large, this becomes,
\beq
	\hat S(k)F({\bf n})=e^{-2ikR}F(-{\bf n}).\label{eigenS}
\eeq
By squaring,
\beq
	\hat S(k)^2F({\bf n})=e^{-4ikR}F({\bf n})\label{eigenS2}
\eeq
Not every solution of (\ref{eigenS2}) is a solution of (\ref{eigenS}). In the limit of large \m{R}, we can take account of this over-counting just dividing  the density of states by two\footnote{For example, in the spherically symmetric case each solution is degenerate, with degeneracy labelled by the angular
momentum quantum numbers. In order to satisfy (\ref{eigenS}), we have
the additional condition that  \m{n} is odd for odd angular momentum \m{l} and
even for even \m{l}. This is not needed for (\ref{eigenS2}) }.

The solutions of \eqn{eigenS2} are the eigenfunctions of the
scattering matrix introduced above; the momenta are fixed by
\beq
	4\eta_{a}(k)+4kR=2\pi n,\quad n=0,1,2,\cdots
\eeq
Thus, the solutions are labelled by \m{n,a} and will depend on \m{R}:
\m{k_{na}(R)}. In the case of the free
particle, the phase shifts vanish and the allowed momentum values  are
\beq
{\pi n\over 2R}.
\eeq

 As \m{R} becomes large the solutions will differ
from this by a small correction:
\beq
	k_{na}(R)={\pi n\over 2R}+{q_{na}(R)\over R}.
\eeq
We get
\beq
	q_{na}(R)=-\eta_a\big({\pi n\over 2 R}\big).
\eeq

Now consider the difference of traces, (remembering to divide by two
to avoid the above mentioned over-counting)
\beq
	\bar{\tr} \phi(\surd H)=\half \sum_{n,a}
\bigg[\phi\big(k_{na}(R)\big)-\phi\bigg(\big\{{\pi n
\over 2 R}\big\}\bigg)\bigg].
\eeq
As  \m{R\to \infty} we get
\beq
	\bar{\tr}\phi(\surd H)=-{1 \over 2R}\sum_{n,a}\phi'(k)
	\eta_a({\pi n \over 2 R})\to -\sum_{a}\int_0^\infty {dk\over \pi}{d \phi(k)\over
dk}\eta_a(k).
\eeq
In other words,
\beq
 \bar{\tr}\phi(\surd H)=-{1\over 2\pi i}\int dk {d\phi(k)\over dk}\tr \log \hat
S(k).
\eeq
We are thinking of \m{\hat S(k)} as  an operator on \m{L^2(S^2)}, so
the trace on the r.h.s. is the average over angles.  We will now turn to a more precise definition of this trace: naively it is logarithmically divergent.

\subsection{A toy model: the Gamma function}
The reader familiar with the theory of modified determinants as in equation (\ref{moddet}) below could skip this subsection.

Recall that the Gamma function has poles at negative integers and zero; its reciprocal is an entire function with zeroes at these points. So we might hope for a product formula
\beq
{1\over \Gamma(z)} \sim z\prod_{n=1}^\infty \left[1+{z\over n}\right]\label{prodGamma}
\eeq
Alas, this product is divergent: the product \m{ \prod_n \left[1+\lambda_n\right]} converges when the sum  \m{\sum_n |\lambda_n|} converges. In the our case, \m{\sum_n{1\over n} } diverges logarithmically. But, the sum of the squares converges: \m{\sum_n{1\over n^2}< \infty}.

This suggests a fix to this divergence problem.The function \m{e^{-z}(1+z)} has the same zero as \m{1+z} but tends to one faster as \m{|z|\to 0}.

Indeed, \m{|e^{-z}(1+z)-1|<C|z|^2}.So the product
\beq
\prod_{n=1}^\infty e^{-{z\over n}}\left[1+{z\over n}\right]
\eeq
converges. So we can separate out a divergent part  of (\ref{prodGamma}) as follows:
\beq
\log{1\over \Gamma(z)}=\log z+\sum_{n=1}^\infty \log\left\{e^{-{z\over n}}\left[1+{z\over n}\right]\right\}+z\sum_{n=1}^\infty {1\over n}
\eeq
The divergence has been isolated to the last term. How to give a meaning to that last divergent sum? We might guess that the correct formula is
\beq
\log{1\over \Gamma(z)}=\log z+\sum_{n=1}^\infty \log\left\{e^{-{z\over n}}\left[1+{z\over n}\right]\right\}+az
\eeq
for some constant \m{a}; and determine it by comparison with the value of the logarithmic derivative of \m{\Gamma(z)} at some point. This is a kind of renormalization  of the logarithmic divergence. Indeed, we know that
\beq
\psi(z)\equiv{d\over dz}\log\Gamma(z)\sim -{1\over z}-\gamma +{\mathrm O}(z)
\eeq
where \m{\gamma} is the Euler constant. Thus we conclude that \m{a=\gamma}:
\beq
{1\over \Gamma(z)}=ze^{\gamma z}\prod_{n=1}^\infty e^{-{z\over n}}\left[1+{z\over n}\right]\label{prodGamma}
\eeq
This argument would not pass muster with a modern analyst. But it is precisely  such  heuristic arguments that led to the  rigorous modern theory of analytic functions. Quantum Field Theory  is still in the stage of development that complex function theory was in the mid nineteenth century: we need to work with heuristic, physically motivated arguments which point the way to the truth. These then should be turned into theorems later, as stronger constructive methods become available.
\subsection{Infinite Determinants}
The determinant of an operator \m{1+X} is well-defined when \m{X}  is trace-class: that is, when the sum of the absolute values of its characteristic values is convergent. But the scattering matrices of interest to us are not this type: \m{\hat S-1} is not  trace-class. However, we will see soon that it has the next best property: the sum of the absolute squares of the characteristic values of \m{\hat S-1} converges, because it is proportional  to the total scattering cross-section. When\m{X} is Hilbert-Schmidt (i.e., \m{\tr X^\dagger X<\infty)}, the modified determinant\cite{BSimon,MickelssonRajeev}
\beq
{\det}_1\left[1+X\right]=\det e^{-X}\left[1+X\right]\label{moddet}
\eeq
is meaningful. (The point is that the analytic function \m{e^{-z}(1+z)-1} is bounded by  \m{c|z|^2} for some constant \m{c}. Hence \m{e^{-X}\left[1+X\right]-1}  is trace-class if \m{X} is Hilbert-Schmidt.  ) Now, we can write the logarithm of the original determinant as
\beq
\log\det[1+X]=\log{\det}_1[1+X]+\tr X
\eeq
which separates out  the divergent term  \m{\tr X}.

{\bf Lemma} \m{\hat S(k)-1} is Hilbert-Schmidt \hfill\break

 {\bf Proof:} \m{\ }In terms if the scattering amplitude,
\beq
	\tr (\hat S(k)-1)^{\dag}(\hat S(k)-1)=4k^2\tr \hat f^\dag f=
		4k^2\int |f(k,{\bf n},{\bf n}')|^2{d\Omega_{\bf
n}d\Omega_{{\bf n}'}\over (4\pi)^2}
\eeq
But the total cross-section for a beam incident along the direction
\m{{\bf n}} is
\beq
	\sigma(k,{\bf n})=\int |f(k,{\bf n},{\bf n}')|^2d\Omega_{{\bf n}'}.
\eeq
Thus
\beq
	\tr (\hat S(k)-1)^{\dag}(\hat S(k)-1)={1\over \pi}  k^{2}\bar \sigma(k)
\eeq
where
\beq
	\bar \sigma(k)={1\over 4\pi}\int d\Omega_{\bf n} \sigma(k,{\bf
n})
\eeq
is the average total cross-section, which is finite as was promised.

Then \m{(1+2ik\hat f)e^{-2ik\hat f}-1} is a trace-class operator and
 the modified determinant \m{{\det}_1\hat S(k)} defined by  \cite{BSimon}
\beq
{\det}_1[\hat S(k)]=\det[(1+2ik\hat f)e^{-2ik\hat f}]
\eeq
exists.  Moreover we have the bound,
\beq
	|\log{\det}_1[\hat S(k)]|<{1\over 2}\tr[ \hat S(k)-1]^{\dag}[
\hat S(k)-1]={1\over 2\pi} k^2\bar\sigma(k)\label{bounddet1S}
\eeq
which will be useful later.

We can write
\beq
 	\log\det\hat S(k)=\log{\det}_1\hat S(k)+2ik\tr \hat f(k).
\eeq
The  real
parts of the terms on the l.h.s must cancel, since the scattering
matrix is unitary.
So we might as well write,
\beq
 	\log\det\hat S(k)=i\Im\log{\det}_1\hat S(k)+2ik\Re\tr \hat f(k).
\eeq
\section{A Dispersion Relation}
The last term, which is potentially divergent,
can be given a meaning using the dispersion relation which relates it to  the scattering
cross-section:(See  \cite{LandauQM})

\beq
	\Re\tr \hat f(k)=-{1\over 2\pi}\int V(x,k)d^3x+{1\over
4\pi^2}{\cal P}\int_0^\infty {2k'^2 \bar \sigma(k')\over k'^2-k^2}dk'.
\eeq
(Here, \m{\cal P} stands for the principal part of the integral.)
The first term is just the Born approximation of the forward
scattering amplitude.
 We will see in a minute that
\beq
	\bar\sigma(k)\sim k^{-2}, \;{\rm as}\; k\to \infty.
\eeq
(This is for the case that the potential \m{V(x)} is independent of \m{k}. It falls off even faster if the potential itself vanishes for large \m{k}.)
Hence this integral is convergent. Thus we have the formula we seek:
\beq
	 \arg\det \hat S(k)=\arg {\det}_1\hat S(k)
		-{k\over \pi}\int V(x,k)d^3x+
{k\over\pi^2}{\cal P}\int_0^\infty {k'^2 \bar \sigma(k')\over k'^2-k^2}dk'.
\eeq
\pagebreak[1]
As in (\ref{trphi1d}) we can now get  trace of a  function of the hamiltonian:
\beqs
	\bar{\tr}\phi\left(\surd H \right)
&=&\int_0^\infty{dk\over 2\pi}{d\phi(k)\over dk}
{k\over \pi} \int V(x,k)d^3x \\
& & -\int_0^\infty{dk\over 2\pi}{d\phi(k)\over dk}{k\over\pi^2}{\cal P}\int_0^\infty {k'^2 \bar \sigma(k')\over k'^2-k^2}dk'\\
& &-\int_0^\infty{dk\over 2\pi}{d\phi(k)\over dk}{\arg\det}_1 S(k)\\
\eeqs
The last term, which is the most complicated to calculate,   can be bounded \eqn{bounddet1S}  by the total cross-section  :
\beq
\left|\int_0^\infty{dk\over 2\pi}{d\phi(k)\over dk}{\arg\det}_1 S(k)\right|<
\int_0^\infty{dk\over 2\pi}{d\phi(k)\over dk}  { k^2 \bar\sigma(k)\over 2\pi}
\eeq

\subsection{Application to Casimir Energy }

An application of the above formula is to the Casimir effect, \m{\phi(k)=k}. To avoid divergences, we assume that \beq
k^2V(x,k)\to 0
\eeq
for large \m{k}. 
\beqs
	\bar{\tr}\surd H  
&=&\int_0^\infty{dk\over 2\pi}
{k\over \pi} \int V(x,k)d^3x \\
& & -\int_0^\infty{dk\over 2\pi}{dk'\over 2\pi }\ \ {kk'\over\pi}\ \  {k' \bar \sigma(k')-k \bar \sigma(k)\over k'^2-k^2}\\
& & -\int_0^\infty{dk\over 2\pi}{\arg\det}_1 \hat{S}(k)
\eeqs
The first two terms should give a good approximation in many cases. Again, the last term can be bounded by the cross-section,
\beq
\left|\int_0^\infty{dk\over 2\pi}{\arg\det}_1 S(k)\right|<
\int_0^\infty{dk\over 2\pi} { k^2 \bar\sigma(k)\over 2\pi}
\eeq

The computation of the scattering matrix from the optical potential is a separate problem: it can be done numerically. Or we could measure the scattering matrix experimentally and use it directly to calculate the Casimir energy.

Despite the physical intuition that the Casimir force is due to reflection of virtual photons,  we do not  get an answer  entirely in terms of the scattering matrix: the first term involves the potential itself. This is similar to the way that the product formula for the Gamma function does not just involve the location of its poles:  our term involving the potential is the analogue of the term involving the Euler constant in the formula for \m{\log\Gamma(z)}. Such `anomalies'  occur elsewhere  in Quantum Field Theory\cite{Anomaly}  as well.

\section*{Acknowledgement}

I thank P. W. Milonni for re-awakening my interest in the Casimir effect;  Arnab Kar for a critical reading; Robert Jaffe for comments on an earlier version of this paper as well as guidance to the literature. This work was supported in part by a grant from the US Department of Energy under contract DE-FG02-91ER40685.

\end{document}